\documentclass[12pt]{iopart}
\begin{document}

\title{Null Lagrangians and Invariance of Action: Exact Gauge Functions}

\author{Z. E. Musielak}
\address{Department of Physics, The University of Texas at 
Arlington, Arlington, TX 76019, USA}
\ead{zmusielak@uta.edu}

\begin{abstract}
Null Lagrangians and their gauge functions are derived for given 
standard and non-standard Lagrangians.  The obtained standard null 
Lagrangians generalize those previously found but the non-standard null 
Lagrangians are new.  The gauge functions are used to make the action
invariant and introduce the exact null Lagrangians, which form a new family 
of null Lagrangians.  The conditions required for the action to be invariant 
are derived for all null Lagrangians presented in this paper, thus, making 
them exact.  As a specific application, the exact null Lagrangians are derived 
for a second-order ordinary differential equation.  The application shows 
that the exact null Lagrangians can be obtained for any ordinary differential 
equation with known Lagrangian.
\end{abstract}

\section{Introduction}

In the calculcus of variations, the action $\mathcal{A} [x(t)]$ is a functional with 
$x(t)$ denoting an ordinary ($x :\ \mathcal{R} \rightarrow \mathcal{R}$) and 
smooth function with at least two continuous derivatives ($\mathcal{C}^{2}$).  
The action $\mathcal{A} [x(t)]$ is defined by an integral over a smooth function 
$L (\dot x, x, t)$ called Lagrangian, where $\dot x =  dx / dt$.  According to the 
principle of least action, or Hamilton's principle [1,2], the action must be 
stationary (to have either a minimum or maximum or saddle point),which is 
mathematically expressed as $\delta \mathcal{A} = 0$, where $\delta$ is the 
functional derivative of the action with respect to $x(t)$.  The necessary condition 
that $L (\dot x, x, t)$ satisfies this principle is $\hat {EL} [ L (\dot x, x, t) ] = 0$, 
where the Euler-Lagrange (EL) operator is 
\begin{equation}
\hat {EL} = {d \over {dt}} \left ( {{\partial} \over {\partial \dot x}} \right ) - 
{{\partial} \over {\partial x}}\ .
\label{S1eq1}
\end{equation}

In general, Lagrangians can be classified as standard, non-standard and null.   
Following the original work by Lagrange [3], standard Lagrangians (SLs) 
depend on $\dot x^2 (t)$ and $x^2 (t)$.  However, non-standard Lagrangians 
(NSLs) do not containt squares of the dependent variable [4].   Null Lagrangians 
are those that solve identically the EL equation and can be expressed as the total 
derivative of a scalar function [5], also called a gauge function [6].  The existence 
of the SLs and NSLs is guaranteed by the Helmholtz conditions [7]; however, the 
existence of null Lagrangians is independent from these conditions.   The process 
of finding a Lagrangian for a given differential equation is called the inverse problem 
of the calculus of variations, or the Helmholtz problem [8,9].  

There are different methods to find the SLs, NSLs and null Lagrangians.  The 
SLs can be found by using either the Jacobi last multiplier [10,11], or by 
fractional derivatives [12], or direct methods [13,14]; there are numerous 
applications of SLs to different physical systems.  Methods to construct the 
NSLs were developed by many authors [13-17].  A novel form of NSLs was 
proposed by El Nabulsi [18], who used a fractional action approach to obtain 
the cosmological equations.  Generalized forms of NSLs were also presented 
[19] and then applied to second-order ordinary differential equations (ODEs), 
whose solutions are special functions of mathematical physics [20].   Other 
generalizations of the NSLs and their applications to nonlinear ODEs were
also made to the Riccati equation [21] and to a Liénard-type nonlinear 
oscillator [22]. 

Null Lagrangians, also known as trivial Lagrangians, have been extensively
investigated in mathematics.  Specific studies involved structures and generation
of these Lagrangians [5,23,24], geometric formulations [25], Cartan and Lepage 
forms and symmetries of Lagrangians [24,26,27], trivial Lagrangians in field 
theories [28,29], and Carath\'eodory's theory of fields of extremals and integral 
invariants [1,24].  Null Lagrangians were also applied to elasticity, where they 
represent the energy density function of materials [30,31], and to demonstrate 
the Galilean invariance of the Newton law of inertia [32].  However, the role of 
null Lagrangians and their gauge functions in physics is not yet established.  

The basis for the Lagrange formalism is the jet-bundle theory and the 
null Lagrangians are typically studied within the frame of this theory 
[28,29].   Let $T$ and $X$ be differentiable manifolds of dimensions 
$m$ and $M+m$, respectively, and let $\pi: X \rightarrow T$ be a 
fibred bundle structure, with $\pi$ being the canonical projection of 
the fibration.  Let $J^r_m (X) \rightarrow T$ be the r-th jet bundle, 
where $t \in T$ and $x \in X$, with $r \in IN$.  Then, an ODE of 
order $q$ is called locally variational (or the E-L type) if, and only if, 
there exists a local real function $L$ constrainted by the condition 
$q \leq r$.  For the ODEs with $q = 2$ the resulting local Lagrangian 
depends on $\dot x (t)$, $x (t)$ and $t$, and it can be written as 
$L (\dot x, x, t)$.   Such local Lagrangians are not unique as other 
Lagrangians may also exist and they would give the same original 
equations when substituted into the E-L equations.  

A general second-order ODE is of the form $\hat D x (t) = 0$, where 
$\hat D = A(t) d \ dt^2 + B(t) d \ dt + C (t)$ is a linear operator and 
$A (t)$, $B (t)$ and $C (t)$ are ordinary and smooth functions to be 
specified.  In order to demonstrate a role that the standard and 
non-standard ENLs may play in the theory of ODEs, the simplest s
econd-order ODE, $\hat D_o x (t) = 0$ with $\hat D_o = d \ dt^2$ 
is considered as an example; the obtained results suggest that 
derivation of the ENLs for $\hat D x (t) = 0$ is straightforward. 

The main objective of this paper is to develop methods to find null 
Lagrangians for given standard and non-standard Lagrangians, and 
introduce a requirement that allows defining a new class of null 
Lagrangians called here the {\it exact null Lagrangians} (ENLs).  
All null Lagrangians derived in this paper are the ENLs and their 
gauge functions are also obtained.  Typically, the SLs and NSLs 
are found for a given ordinary or partial differential equation by 
solving the inverse (or Helmholtz) problem of the calculus of 
variations.  In this paper, our approach is different, namely, we 
specify forms of the SLs and NSLs that have been commonly 
used and then present methods of finding exact null Lagrangians 
for the given SLs and NSLs.  Since the forms of the SLs and NSLs 
are very different, the resulting standard and non-standard ENLs 
are also distinct.   

The outline of the paper is as follows: in Section 2, the standard 
and non-standard null Lagrangians and their gauge functions 
are constructed; in Section 3, invariance of the action is described 
and used to define the exact null Lagrangians; application of the 
obtained results to a simple second-order ordinary differential 
equation is presented in Section 4; and the conclusions are 
given in 5.

\section{Null Lagrangians and their gauge functions}

\subsection{Standard null Lagrangians}

The standard Lagrangians $L_s [\dot x (t), x(t)]$ considered here are those 
that depend on the squares of the dependent variable $x^2 (t)$, and its 
derivative $\dot x^2 (t)$, and they are of the form  
\begin{equation}
L_s [\dot x (t), x(t)] = {1 \over 2} [ \alpha \dot x^2 (t) - \beta x^2 (t) ]\ ,
\label{S2eq1}
\end{equation}
where $\alpha$ and $\beta$ are either functions of the independent variable 
or they are constants; in both cases, they must be determined once a 
mathematical or physical problem is specified; in the original Lagrange 
formulation [3], $\alpha = \alpha_o$ = const and $\beta = \beta_o$ 
= const.   

In the previous work [5,23,24,32], null Lagrangians were constructed 
for a given differential equation.  However, here we propose a method to 
construct null Lagrangians for the SL given by Eq. (\ref{S2eq1}).  The basis
of this method is the requirement that the constructed null Lagrangian contains
only terms with the dependent or independent variables or their combination,
which are of lower oder than those present in $L_s [\dot x (t), x(t)]$.  Following 
\cite{32}, the following null Lagrangian is obtained
\begin{equation}
L_{s,n} [\dot x (t), x(t), t] = c_1 \dot x (t) x (t) + c_2 [ \dot x (t) t + x (t) ] 
+ c_3 \dot x (t) + c_4\ ,
\label{S2eq2}
\end{equation}
and its gauge function
\begin{equation}
\Phi_{s,n} [x(t), t] = {1 \over 2} c_1 x^2 (t) + c_2 x (t) t + c_4 x + c_6 t\ ,
\label{S2eq3}
\end{equation}
where the subscript 'n' stands for null, and $c_1$, $c_2$, $c_3$ and $c_4$ 
are constants to be determined that can be expressed in terms of $\alpha$ 
and $\beta$.  In the previous work, $L_{s,n} [\dot x (t), x(t), t]$ was 
obtained first and then its gauge function $\Phi_n [x(t), t]$ was derived.  
Here, we generalize these results by replacing the constants in either 
$L_{s,n} [\dot x (t), x(t), t]$ or $\Phi_{s,n} [x(t), t]$ by ordinary and 
smooth functions.  The resulting null Lagrangians and their gauge 
functions are presented in Propositions 1 and 2.  

{\bf Proposition 1:}
Let $f_1 (t)$, $f_2 (t)$, $f_3 (t)$ and $f_4 (t)$ be ordinary and smooth 
functions of the following test-Lagrangian
\begin{equation}
L_{s,t} [\dot x (t), x(t), t] = f_1 (t) \dot x (t) x (t) + f_2 (t) [ \dot x (t) t + 
x (t) ] + f_3 (t) \dot x (t) + f_4 (t)\ .
\label{S2eq4}
\end{equation}
Then, $L_{s,t} [\dot x (t), x(t), t]$ is a null Lagrangian if, and only if, $f_1 (t) = 
c_1$, $f_2 (t) = c_2$ and $f_3 (t) = c_3$, where $c_1$, $c_2$ and $c_3$
are constants, and $f_4 (t)$ is arbitrary.

{\bf Proof:}
The necessary condition that $L_{s,t} [\dot x (t), x(t), t]$ is a null Lagrangian 
is $\hat {EL} \{ L_{s,t} [\dot x (t), x(t), t] \} = 0$, which gives 
\begin{equation}
\dot f_1 (t) x (t) + \dot f_2 (t) t + \dot f_4 (t) = 0\ . 
\label{S2eq5}
\end{equation}
This condition is only satisfied when $f_1 (t) = c_1$, $f_2 (t) = c_2$ and 
$f_3 (t) = c_3$, and it does not depend on $f_4 (t)$, which implies that 
this function can be arbitrary.  

Comparison of $L_{s,t} [\dot x (t), x(t), t]$ obtained in Proposition 1 to 
$L_{s,n} [\dot x (t), x(t), t]$ of Eq. (\ref{S2eq2}) can be summarized as 
follows. 

{\bf Corollary 1:}
Any arbitrary constant or arbitrary function of independent variable may 
be added to any null Lagrangian without changing the properties of this 
Lagrangian.

{\bf Proposition 2:}
Let the general null Lagrangian be 
\[
L_{s,gn} [\dot x(t), x(t), t] = [ f_1 (t) \dot x (t) + {1 \over 2}
\dot f_1 (t) x (t) ] x (t) + \left [ f_2 (t) \dot x (t) + \dot f_2 
(t) x (t) \right ] t 
\]
\begin{equation}
\hskip0.25in + f_2 (t) x (t)  + \left [ f_3 (t) \dot x (t) + 
\dot f_3 (t) x (t) \right ] + \left [ f_4 (t) + \dot f_4 (t) \right ]\ .
\label{S2eq6}
\end{equation}
where $f_1 (t)$, $f_2 (t)$, $f_3 (t)$ and $f_4 (t)$ are ordinary 
and smooth functions of the independent variable.  Then, the 
null Lagrangian $L_{s,gn} [\dot x(t), x(t), t]$ is obtained if, 
and only if, the gauge function $\Phi_{s,n} [x(t), t]$ (see Eq. 
\ref{S2eq3}) is generalized to 
\begin{equation}
\Phi_{s,gn} [x(t), t] = {1 \over 2} f_1 (t) x^2 (t) + f_2 (t) 
x (t) t + f_3 (t) x (t) + f_4 (t) t\ ,
\label{S2eq7}
\end{equation}
which becomes the general gauge function.

{\bf Proof:}
In order for $L_{s,gn} [\dot x(t), x(t), t]$ to be the general null
Lagrangian, it is required that $\hat {EL} \{ L_{s,gn} [\dot x (t), 
x(t), t] \} = 0$ and 
\begin{equation}
L_{s,gn} [\dot x(t), x(t), t] = {{d \Phi_{s,gn} [ x (t), t ]} \over 
{dt}} = {{\partial \Phi_{s,gn} [ x (t), t ]} \over {\partial t}} +
\dot x (t) {{\partial \Phi_{s,gn} [ x (t), t ]} \over {\partial x}}\ . 
\label{S2eq8}
\end{equation}
Substitution of Eq. (\ref{S2eq7}) into Eq. (\ref{S2eq8}) gives 
the general null Lagrangian.

The following Corollary extends (without a proof) the results 
of Proposition 2 to all null Lagrangians of this category. 

{\bf Corollary 2:}
The Lagrangian $L_{s,gn} [\dot x(t), x(t), t]$ is the most general 
null Lagrangian among all null Lagrangians that can be constructed 
from terms with the lowest orders of the dependent variable and its 
derivative.

{\bf Corollary 3:}
{For the special cases of $f_1 (t) = c_1$, $f_2 (t) = c_2$, $f_3 (t) = c_3$
and $f_4 (t) = c_4$, the general null Lagrangian $L_{s,gn} [\dot x(t), 
x(t), t]$ reduces to that given by Eq. (\ref{S2eq2}) or the one presented 
in [32].

To determine the arbitrary functions in the standard null Lagrangians,
additional conditions are necessary; such conditions are introduced in
Section \ref{sec3}, where it is shown how the coefficients $\alpha$
and $\beta$ in the standard Lagrangian $L_s [\dot x (t), x(t)]$ are 
related to the arbitrary functions in $L_{s,gn} [\dot x (t), x(t), t]$.

\subsection{Non-standard null Lagrangians}

Any Lagrangian different than $L_s [\dot x (t), x(t)]$ (see Eq. 
\ref{S2eq1}) is called a non-standard Lagrangian (NSL).  
Among known NSLs, the most commonly used is 
\begin{equation}
L_{ns} [\dot x(t), x(t), t] = {{1} \over {g_1 (t) \dot x (t) +
g_2 (t) x (t) + g_3 (t)}}\ ,
\label{S2eq9}
\end{equation}
where $g_1 (t)$, $g_2 (t)$ and $g_3 (t)$ are ordinary and smooth
functions to be determined in such a way that a second-order ODE 
derived by using $L_{s} [\dot x(t), x(t), t]$ is identical to that obtained 
when $L_{ns} [\dot x(t), x(t), t]$ is substituted into the EL equation. 
As shown by Eq. (\ref{S2eq1}), this derivation of the ODE depends
on the form of the coefficients $\alpha$ and $\beta$, and whether 
these coefficients are constants or functions of the independent 
variable.  In other words, a prior knowledge of the ODE resulting 
from the standard Lagrangian is needed in order to determine the 
forms of the functions $g_1 (t)$, $g_2 (t)$ and $g_3 (t)$ [13,14,
16,17]; see Section 4 for details.

However, for the purpose of finding non-standard null Lagrangians,
determination of $g_1 (t)$, $g_2 (t)$ and $g_3 (t)$ is not required.
Since there are no non-standard null Lagrangians published in the 
literature, in the following we present our first such Lagrangians 
and demonstrate their association with $L_{ns} [\dot x(t), x(t), t]$.  
To find these Lagrangians, we impose two conditions.  First, a 
form of any non-standard null Lagrangian must be similar to that 
of $L_{ns} [\dot x(t), x(t), t]$, and second, the order of the dependent
variable and its derivative must not exceed their order in the NSL 
given by Eq. (\ref{S2eq9}).  The obtained non-standard null 
Lagrangians are presented in Propositions 3 and 4, and the 
Corollaries that follow them. 

{\bf Proposition 3:}
Let $a_1$, $a_2$, $a_3$ and $a_4$ be constants in the following 
non-standard test-Lagrangian
\begin{equation}
L_{ns,t} [\dot x (t), x(t), t] = {{a_1 \dot x (t)} \over {a_2 x (t) + a_3 t 
+ a_4}}\ .
\label{S2eq10}
\end{equation}
Then, $L_{ns,t} [\dot x (t), x(t), t]$ is a null Lagrangian if, and only if, 
$a_3 = 0$.

Substitution of this Lagrangain into the EL equation, $\hat {EL} \{ L_{ns,t} 
[\dot x (t), x(t), t] \} = 0$ that gives $a_1 a_3 = 0$; since $a_1 \neq 0$,
then $a_3 = 0$.

{\bf Corollary 4:}
Let $L_{ns,n} [\dot x (t), x(t)]$ be the non-standard null Lagrangian with 
$a_3 = 0$.  then, its gauge function is given by $\Phi_{ns,n} [x (t)] = 
(a_1 / a_2) \ln \vert a_2 x (t) + a_4 \vert$.

{\bf Corollary 5:}
Another null Lagrangian that can be constructed is $L_{tn} (t) = b_1 / 
( b_2 t + b_3 )$ with its gauge function $\Phi_{tn} (t) = (b_1 / b_2) 
\ln \vert b_2 t + b_3 \vert$; however, this Lagrangian and its gauge 
function do not obey the first condition, thus, they will not be further 
considered. 

Our generalization of the gauge function $\Phi_{ns,n} (t)$ given in 
Proposition 3 is now presented in the following proposition.

{\bf Proposition 4:}
Let $h_1 (t)$, $h_2 (t)$ and $h_4 (t)$ be ordinary and smooth 
functions in the following general, non-standard, test-gauge function
\begin{equation}
\Phi_{ns,gt} [\dot x (t), x(t), t] = \left [ {{h_1 (t)} \over {h_2 (t)}} 
\right ] \ln [ h_2 (t) x (t) + h_4 (t)]\ .
\label{S2eq11}
\end{equation}
This gauge function gives a null Lagrangian if, and only if, 
$h_4 (t) = a_4$ = const.

{\bf Proof:}
Using Eq. (\ref{S2eq8}), the gauge function $P_{ns,gt} [\dot x (t), 
x(t), t]$ gives the general non-standard test-Lagrangian 
\[
L_{ns,gt} [\dot x (t), x(t), t] = {{h_1 (t) \dot x (t)} \over {h_2 (t) 
x (t) + h_4 (t)}} + {{h_1 (t)} \over {h_2 (t)}} {{\dot h_2 (t) x (t) 
+ \dot h_4 (t)} \over {h_2 (t) x (t) + h_4 (t)}}
\]
\begin{equation}
\hskip0.25in + \left [ {{\dot h_1 (t)} \over {h_2 (t)}} - {{\dot h_1 
(t) \dot h_2 (t)} \over {h^2_2 (t)}} \right ]\ \ln \vert h_2 (t) x (t)
+ h_4 (t) \vert\ .
\label{S2eq12}
\end{equation}
Verification of $\hat {EL} \{ L_{ns,gt} [\dot x (t), x(t), t] \} = 0$ 
leads to
\begin{equation}
{{\dot h_4 (t)} \over {h_2 (t) x (t) + h_4 (t)}} = 0\ ,
\label{S2eq13}
\end{equation}
which shows that $\dot h_4 (t) = 0$ or $h_4 (t) = a_4$ = const.  Thus, 
the final form of the general, non-standard, null Lagrangian is 
\[
L_{ns,gn} [\dot x (t), x(t), t] = {{h_1 (t) \dot x (t)} \over {h_2 (t) 
x (t) + a_4}} + {{h_1 (t)} \over {h_2 (t)}} {{\dot h_2 (t) x (t)} 
\over {h_2 (t) x (t) + a_4}}
\]
\begin{equation}
\hskip0.25in + \left [ {{\dot h_1 (t)} \over {h_2 (t)}} - {{\dot h_1 
(t) \dot h_2 (t)} \over {h^2_2 (t)}} \right ]\ \ln \vert h_2 (t) x (t)
+ a_4 \vert\ ,
\label{S2eq14}
\end{equation}
and this is the general non-standard null Lagrangian, whose gauge 
function is given by Eq. (\ref{S2eq11}).

According to the results of Proposition 4, the general, non-standard gauge 
function for $L_{ns,gn} [\dot x (t), x(t), t]$ is 
\begin{equation}
\Phi_{ns,gn} [\dot x (t), x(t), t] = \left [ {{h_1 (t)} \over {h_2 (t)}} 
\right ] \ln [ h_2 (t) x (t) + a_4 ]\ .
\label{S2eq15}
\end{equation}

{\bf Corollary 6:}
The Lagrangian $L_{ns,gn} [\dot x (t), x(t), t]$ reduces to 
$L_{ns,n} [\dot x (t), x(t), t]$, and the gauge function $\Phi_{ns,gn} 
[\dot x (t), x(t), t]$ becomes $\Phi_{ns,n} [\dot x (t), x(t), t]$ when
the functions $h_1 (t)$ and $h_2 (t)$ are replaced by the constants
$a_1$ and $a_2$, respectively.

The derived $L_{ns,gn} [\dot x (t), x(t), t]$ and its gauge function 
represent the most general non-standard null Lagrangian and the 
gauge function that can be obtained for the NSL given by Eq. 
(\ref{S2eq9}).  This is based on the condition that the order of 
the dependent variable in the null Lagrangian is either the same 
or lower than that displayed in the NSL.

\section{Invariance of action and exact null Lagrangians}

Our main results derived in the previous section are the general, 
standard, null Lagrangian given by Eq. (\ref{S2eq6}), and the 
general, non-standard, null Lagrangian given by Eq. (\ref{S2eq12}),
and their gauge functions.  The Lagrangians and the gauge functions 
depend on arbitrary functions that must be ordinary and smooth.  
These functions must obey certain constraints, which are now 
specified.   

In general, the action is defined as   
\[
A [x (t); t_e, t_o] = \int^{t_e}_{t_o} ( L + L_{\rm null} ) dt = 
\int^{t_e}_{t_o} L dt + \int^{t_e}_{t_o} \left [ {{d \Phi_{\rm null} 
(t)} \over {dt}} \right ] dt 
\]
\begin{equation}
\hskip0.25in = \int^{t_e}_{t_o} L dt + [ \Phi_{\rm null} (t_e) - 
\Phi_{\rm null} (t_o)]\ , 
\label{S3eq1}
\end{equation}
where $t_o$ and $t_e$ denote the initial and final times, $L$ is a 
Lagrangian that can be either any SL or any NSL, $L_{\rm null}$
is any null Lagrangian and $\Phi_{\rm null}$ is its gauge function.
Since both $\Phi_{\rm null} (t_e)$ and $\Phi_{\rm null} (t_o)$ are 
constants, they do not affect the Hamilton Principle that requires 
$\delta A [x (t)] = 0$.  However, the requirement that $\Delta 
\Phi_{\rm null} = \Phi_{\rm null} (t_e) - \Phi_{\rm null} (t_o)$
= const adds this constant to the value of the action.  In other 
word, the value of the action is affected by the gauge function.

Let us now define a new category of null Lagrangians, which have 
one additional characteristic, namely, they make the action gauge 
invariant, or more precisely independent from the gauge function.

{\bf Definition 1:}
A null Lagrangian whose $\Delta \Phi_{\rm null} = 0$ at the end 
conditions is called the exact null Lagrangian.

{\bf Definition 2:}
A gauge function with properties that $\Delta \Phi_{\rm null} = 0$
is called the exact gauge function.

The condition $\Delta \Phi_{\rm null} = 0$ is satisfied when 
$\Phi_{\rm null} (t_e) = 0$ and $\Phi_{\rm null} (t_o) = 0$.  Thus,
the exact null Lagrangians (ENLs) are those whose exact gauge functions 
(EGFs) are zero at the end conditions.

We use these conditions to establish constraints on arbitrary functions 
that the null Lagrangians and their gauge functions depend on.  We 
begin with $L_{s,gn} [\dot x(t), x(t), t]$ (see Eq. \ref{S2eq6}) 
and its gauge function $\Phi_{s,gn} [x(t), t] $ (see Eq. \ref{S2eq7}),
and this Lagrangian is exact if, and only if, $\Phi_{s,gn} (t_e) = 0$ 
and $\Phi_{s,gn} (t_o) = 0$, or explicitly
\begin{equation}
{1 \over 2} f_1 (t_e) x_e^2 + f_2 (t_e) x_e t_e + f_3 (t_e) x_e + 
f_4 (t_e) t_e = 0\ ,
\label{S3eq2}
\end{equation}
and 
\begin{equation}
{1 \over 2} f_1 (t_o) x_o^2 + f_2 (t_o) x_o t_o + f_3 (t_o) x_o + 
f_4 (t_o) t_o = 0\ ,
\label{S3eq3}
\end{equation}
with $x_e = x (t_e)$ and  $x_o = x (t_o)$ at the end points.  If the  
arbitrary functions satisfy these conditions, then the action remains 
invariant, and $L_{s,gn} [\dot x(t), x(t), t]$ is the general, standard, 
exact null Lagrangian.  It is seen that the relationships $f_3 (t_e) = 
- f_1 (t_e) x_e / 2$ and $f_4 (t_e) = - f_2 (t_e) x_e$ solve the first 
condition, and similar relationships for $t_o$ solve the second condition; 
this shows how these functions can be related to each other.

Applying the same procedure to $L_{ns,gn} [\dot x (t), x(t), t]$ and
its gauge function $\Phi_{ns,gn} [\dot x (t), x(t), t]$, we obtain
\begin{equation}
\left [ {{h_1 (t_e)} \over {h_2 (t_e)}} \right ] \ln [ h_2 (t_e) x_e + 
a_4] = 0\ ,
\label{S3eq4}
\end{equation}
and 
\begin{equation}
\left [ {{h_1 (t_o)} \over {h_2 (t_o)}} \right ] \ln [ h_2 (t_o) x_o + 
a_4] = 0\ .
\label{S3eq5}
\end{equation}
Since $\ln [ h_2 (t_e) x (t) + a_4] \neq 0$, these conditions set up 
stringent limits on the function $h_1 (t)$, whose end values must be 
$h_1 (t_e) = 0$ and $h_1 (t_o) = 0$; however, this procedure does 
not impose any constraint on $h_2 (t)$.  

Further constraints on all arbitrary functions that appear in the general 
standard and non-standard ENLs can be imposed by considering 
symmetries of these Lagrangians and the resulting dynamical equations.   
In general, Lagrangians posses less symmetry than the equations they 
generate [33].  Among different symmetries, Noether and non-Noether 
symmetries are identified [34-36].  The presence of null Lagrangians 
does not affect the Noether symmetries [33,36]; however, it may have 
effects on the non-Noether symmetries [32,37].  Studies of these 
symmetries will give new constraints on the functions and they will 
be investigated elsewhere. 

\section{Application of exact null Lagrangians}

To demonstrate applications of the ENLs to ODEs, we select the simplest 
second-order ODE, and present the general standard and non-standard 
ENLs for this equation.  The selection of this ODE has an important 
advantage because it represents the first Newtonian law of dynamics,
also know as the {\it law of inertia}, which requires that motion of 
any classical body is always rectilinear and uniform with respect to 
any inertial frame of reference.  Let $(x,y,z)$ be a Cartesian coordinate 
system and $t$ be the same time in all inertial frames, then the 
one-dimensional motion of the body in one inertial frame is given by 
$\hat D_o x (t) = 0$ with the end conditions $t_o = 0$ and $t_e = 1$, 
and with the initial conditions: $x (0) = x_o = 1$, $x (1) = x_1 = 2$, 
and $\dot x (0) = u_o$.  Then, the solution to the equation is $x (t) = 
u_o t + 1$.  

In the following, we derive standard and non-standard exact null 
Lagrangians and their corresponding exact gauge functions, and the 
obtained results will be valid for for both the simplest second-order 
ODE and the law of inertia.  Our results significantly generalize the 
standard null Lagrangian derived in [32]; however, the obtained
non-standard exact null Lagrangians are presented for the first 
time in the literature.  

Taking $\alpha = 1$, the SL (see Eq. \ref{S2eq1}) for $\hat D_o x (t) 
= 0$ is 
\begin{equation}
L_s [\dot x (t), x(t)] = {1 \over 2} \dot x^2 (t)\ ,
\label{S4eq1}
\end{equation}
and the general standard ENL is given by Eq. (\ref{S2eq6}) with its 
gauge function
\begin{equation}
\Phi_{s,gn} [x(t), t] = {1 \over 2} f_1 (t) x^2 (t) + f_2 (t) 
x (t) t + f_3 (t) x + f_4 (t) t\ .
\label{S4eq2}
\end{equation}
For the null Lagrangian to be exact, the action must be invariant,
which requires that $\Phi_{s,gn} (1) = 0$ and $\Phi_{s,gn} (0) 
= 0$.  Using these conditions, the following relationships (see 
Eqs. \ref{S3eq2} and \ref{S3eq3}) between the arbitrary 
functions are obtained
\begin{equation}
f_1 (1) + f_2 (1) + f_3 (1) + f_4 (1) = 0\ ,
\label{S4eq3}
\end{equation}
and 
\begin{equation}
f_3 (0) = - {1 \over 2} f_1 (0)\ .
\label{S4eq4}
\end{equation}
If the functions satisfy these relatiosnhips, then $L_{s,gn} [\dot x(t), x(t), t]$
is the ENL, and the gauge function $\Phi_{s,gn} [x(t), t]$ is the EGF.

Let us now use Eq. (\ref{S2eq9}) to obtain the non-standard Lagrangian for
$\hat D_o x (t) = 0$ with the initial conditions specified above.  Since the 
functions $g_1 (t)$, $g_2 (t)$ and $g_3 (t)$ depend on the form of the 
considered ODE, they must be independently derived for each ODE.  General
conditions on the functions in $L_{ns} [\dot x(t), x(t), t]$ derived in [16,19]
can be simplified to 
\begin{equation}
{{g_2 (t)} \over {g_1 (t)}} + {1 \over 3} {{\dot g_1 (t)} \over {g_1 (t)}} 
= 0\ ,
\label{S4eq5}
\end{equation}
\begin{equation}
{{\dot g_2 (t)} \over {g_1 (t)}} - {1 \over 2} {{\dot g_1 (t)} \over {g_1 (t)}}
{{g_2 (t)} \over {g_1 (t)}} + {{g^2_2 (t)} \over {2 g^2_1 (t)}} = 0\ ,
\label{S4eq6}
\end{equation}
and 
\begin{equation}
{{\dot g_3 (t)} \over {g_1 (t)}} - {1 \over 2} {{\dot g_1 (t)} \over {g_1 (t)}}
{{g_3 (t)} \over {g_1 (t)}} + {{g_3 (t)} \over {g_1 (t)}} {{g_2 (t)} \over 
{2 g_1 (t)}}= 0\ .
\label{S4eq7}
\end{equation}
Eliminating $g_2 (t)$ from Eqs. (\ref{S4eq5}) and (\ref{S4eq6}), and 
defining $u (t) = \dot g_1 (t) / g_1 (t)$, we obtain 
\begin{equation}
\dot u (t) + {1 \over 3} u^2 (t) = 0\ ,
\label{S4eq8}
\end{equation}
which is a special form of the Riccati equation. Following [20], the 
solution to Eq. (\ref{S4eq8}) is 
\begin{equation}
u (t) = 3 {{\dot v (t)} \over {v (t)}}\ ,
\label{S4eq9}
\end{equation}
with $v(t)$ representing a solution to $\ddot v (t) = 0$ which is the 
auxiliary condition [19,20].  Using the initial conditions $v (t=0) 
= v_o$ and $\dot v (t=0) = a_o$, the solution becomes $v (t) = 
a_o t + v_o$, and it gives 
\begin{equation}
g_1 (t) = C_1 ( a_o t + v_o)^3\ ,
\label{S4eq9}
\end{equation}
where $C_1$ is an integration constant.  Having obtained $g_1 (t)$, 
we get
\begin{equation}
g_2 (t) = - C_1 a_o ( a_o t + v_o)^2\ .
\label{S4eq10}
\end{equation}
Finally, $g_3 (t)$ can be found by eliminating $g_1 (t)$ and $g_2 (t)$
from Eq. (\ref{S4eq7}) and solving it.  The solution is 
\begin{equation}
g_3 (t) = C_1 C_2 ( a_o t + v_o)^2\ ,
\label{S4eq11}
\end{equation}
where $C_2$ is an integration constant.  

The final form of the non-standard Lagrangian for $\hat D_o x (t) = 0$
is 
\begin{equation}
L_{ns} [\dot x(t), x(t), t] = {{1} \over {C_1 ( a_o t + v_o)^2}}\
{{1} \over {( a_o t + v_o) \dot x (t) - a_o x (t) + C_2}}\ .
\label{S4e12}
\end{equation}
Despite its complexity, as compared to the standard Lagrangian given 
by Eq. (\ref{S4eq1}), $L_{ns} [\dot x(t), x(t), t]$ gives $\hat D_o 
x (t) = 0$ when substituted into the EL equation. 

The general non-standard null Lagrangian (Eq. \ref{S2eq14}) 
and its general gauge function (Eq. \ref{S2eq15}) become 
exact if, and only if, the action remains invariant.  To assure the
invariance of the action, the conditions given by Eqs. (\ref{S3eq4})
and (\ref{S3eq5}) must be satisfied.  This requires that $h_1 (1) =
0$ and $h_1 (0) = 0$; however, no contraints apply to the function
$h_2 (t)$, which can be of any form as long as it is ordinary and 
smooth.

\section{Conclusions}

General null Lagrangians of the lowest order of the dependent variable and 
its derivative are derived for given standard and non-standard Lagrangians.  
The standard null Lagrangians represent a generalization of those previously
found; however, the presented non-standard null Lagrangians are new.  For 
all null Lagrangians, their gauge functions are also obtained.  The gauge 
functions are used to make the action invariant, which allows introducing
a new family of exact null Lagrangians.  The conditions required for the 
action to be invariant are established for all derived null Lagrangians.  
As a result, all presented null Lagrangians are exact.  

The obtained results are used to derive the exact null Lagrangians for the 
simplest second-order ODE, which also is the main ODE of the Newtonian
law of inertia.  The derived standard exact null Lagrangians and their 
corresponding exact gauge functions significantly generalize the previously 
obtained results [32].  However, all non-standard exact null Lagrangians 
are new.   Comparison between the standard and non-standard exact null 
Lagrangians shows significant differences between them.  The presented 
applications demonstrate that the exact null Lagrangians can be obtained 
for any ODE whose standard and non-standard Lagrangians are known, 
and the results would also be valid for any law of physics represented by
this equation.

Finally, let us briefly summarize important open problems resulting from 
this paper.  The presented methods of finding standard and non-standard 
exact null Lagrangians and their exact gauge functions can be extended 
to other ODEs both homogeneous and inhomogeneous. A possible 
extension of this approach to partial differential equations (PDEs) would 
make it applicable to many problems of applied mathematics and theoretical 
physics.  Since there are other families of non-standard Lagrangians, it would 
be interesting to explore methods of finding exact null Lagrangians for each 
family.  More constraints on the arbitrary functions in the exact null Lagrangians 
and their exact gauge functions are necessary, and they can be obtained by 
investigating symmetries and their underlying groups, thus, Lie groups 
associated with the derived exact null Lagrangians, and their exact gauge
functions must be determined [38].

\bigskip\noindent
{\bf Acknowledgments}
This work was supported by the Alexander von Humboldt 
Foundation (Z.E.M.)
%


\begin{thebibliography}{10}


\bibitem{1} M. Giaquinta, S. Hilderbrandt, {\it Calculus of Variations I}
                  (Springer, Berlin, 1996).
\bibitem{2} N.A. Doughty, {\it Lagrangian Interaction} (Addison-Wesley, 
                  New York, NY, USA, 1990). 
\bibitem{3} J.L Lagrange, {\it Analytical Mechanics} (Springer, Netherlands, 
                 1997). 
\bibitem{4} V.I. Arnold, {\it Mathematical Methods of Classical Mechanics} 
                 (Springer, New York, NY, USA, 1978).
\bibitem{5} P.J. Olver, {\it Applications of Lie Groups to Differential Equations} 
                  (Springer, New York,  NY, USA, 1993).
\bibitem{6} J.-M. Levy-Leblond, "Group-theoretical foundations of classical mechanics: 
                  the Lagrangian gauge problem", {\it Commun. Math. Phys.}, 12, 64--79, 
                  1969).
\bibitem{7} H. Helmholtz, "On the physical meaning of the principle of least action", 
                  {\it J.  Reine Angew Math.}, 100, 213, 1887.
\bibitem{8} J. Douglas, "Solution of the inverse problem of the calculus of variations", 
                 {\it Trans. Am. Math. Soc.}, 50, 71--128, 1941.
\bibitem{9} Lopuszanski, J., {\it The Inverse Variational Problems in Classical Mechanics}
                 (World Scientific, Singapore, 1999).
\bibitem{10} M.C. Nucci and P.G.L. Leach, "Lagrangians galore", {\it J. Math. Phys.}, 48, 
                   123510, 2007. 
\bibitem{11} A.G. Choudhury, P. Guha and B. Khanra, "On the Jacobi last multiplier, 
                    integrating factors and the Lagrangian formulation of differential equations 
                    of the Painlevé–Gambier classification", {\it J. Math.Anal. Appl.}, 360,
                    651--664, 2009.
\bibitem{12} F. Riewe, "Nonconservative Lagrangian and Hamiltonian mechanics", 
                   {\it Phys. Rev. E}, 53, 1890, 1996.
\bibitem{13} Z.E. Musielak, "Standard and non-standard Lagrangians for dissipative 
                   dynamical systems with variable coefficients", {\it J. Phys. A Math. Theor.}, 
                   41, 055205, 2008.
\bibitem{14} J.L. Cie\'sli\'nski and T. Nikiciuk, "A direct approach to the construction of 
                   standard and non-standard Lagrangians for dissipative-like dynamical systems 
                   with variable coefficients", {\it J. Phys. A Math. Theor.}, 43, 175205, 2010. 
\bibitem{15} A.I. Alekseev and B.A. Arbuzov, "Classical Yang-Mills field theory with 
                   nonstandard Lagrangians", {\it Theor. Math. Phys.}, 59, 372--378, 1984.
\bibitem{16} Z.E. Musielak, "General conditions for the existence of non-standard Lagrangians 
                   for dissipative dynamical systems", {\it Chaos, Solitons  Fractals}, 42, 2640, 2009.
\bibitem{17} A. Saha and B. Talukdar, "Inverse variational problem for nonstandard Lagrangians",
                   {\it Rep. Math. Phys.}, 73, 299--309, 2014. 
\bibitem{18} R.A. El-Nabulsi, "Fractional action cosmology with variable order parameter",  
                   {\it Int. J. Theor. Phys.}, 56, 1159, 2017.
\bibitem{19} N. Davachi and Z.E. Musielak, "Generalized non-standard Lagrangians",
                   {\it J. Undergrad. Rep. Phys.}, 29, 100004, 2019.
\bibitem{20} Z.E. Musielak, N. Davachi and M. Rosario-Franco, "Special Functions of
                    Mathematical Physics: A Unified Lagrangian Formalism", {\it Mathematics}, 
                    8, 379, 2020.
\bibitem{21} J.F. Carinena, M.F. Ranada and M. Santander, "Lagrangian formalism for nonlinear 
                   second-order Riccati systems: one-dimensional integrability and two-dimensional 
                   superintegrability", {\it J. Math. Phys.}, 46, 062703, 2005.
\bibitem{22} V.K. Chandrasekar, M. Senthilvelan and M. Lakshmanan, "Unusual Liénard-type 
                    nonlinear oscillator", {\it Phys. Rev. E}, 272, 066203, 2005. 
\bibitem{23} P.J. Olver and J. Sivaloganathan, "The structure of null Lagrangians", 
                   {\it Nonlinearity}, 1, 389-398, 1989.
\bibitem{24} M. Crampin and D.J. Saunders, "On null Lagrangians", {\it Diff. Geom. Appl.}, 
                   22, 131--146, 2005. 
\bibitem{25} R. Vitolo, "On different geometric formulations of Lagrange formalism", {\it Diff. 
                   Geom. Appl.}, 10, 293-305, 1999.
\bibitem{26} D.E. Betounes, "Differential geometric aspects of the Cartan form: Symmetry theory",
                   {\it J. Math. Phys.}, 28, 2347, 1987.
\bibitem{27} D. Krupka, O. Krupkova and D. Saunders, "The Cartan form and its generalizations 
                    in the calculus of variations", {\it Int. J. Geom. Meth. Mod. Phys.}, 7, 631 -- 654,
                    2010. 
\bibitem{28} D.R. Grigore, "Trivial second-order Lagrangians in classical field theory", {\it J. Phys. 
                   A}, 28, 2921, 1995.
\bibitem{29} D. Krupka and J. Musilova, "Trivial Lagrangians in field theory", {\it Diff. Geom. Appl.}, 
                   9, 225, 1998.
\bibitem{30} D.R. Anderson, D.E. Carlson and J. Fried, "A continuum-mechanical theory for 
                   nematic elastomers", {\it Elasticity}, 56, 35 -- 58, 1999.
\bibitem{31} G. Saccomandi and R. Vitolo, "Null Lagrangians for nematic elastomers",
                    {\it J. Math. Sciences}, 136, 4470 -- 4477, 2006.
\bibitem{32} Z.E. Musielak and T.B. Watson, "Gauge functions and Galilean invariance of 
                   Lagrangians", {\it Phys. Let. A}, 384, 126642, 2020.
\bibitem{33} S. Hojman, "Symmetries of Lagrangians and of their equations of motion",
                   {\it J. Phys. A: Math. Gen.} 17, 2399--2412, 1984.
\bibitem{34} A.K. Halder, A. Palithanasis and P.G.L. Leach, "Noether's theorem and 
                   symmetries", {\it Symmetry} 10, 744 (21 pages), 2018.
\bibitem{35} G.F. Torres del Castillo, "Point symmetries of the Euler-Lagrange equations",
                   {\it Rev. Mex. Fisica} 60, 129 --135, 2014.
\bibitem{36} W. Sarlet, "Note on equivalent Lagrangians and symmetries", {\it J. Phys. A: Math. Gen.}, 
                   16, L229 -- L233, 1983.
\bibitem{37} S.A. Hojman, "A new conservation law constructed without using either Lagrangians 
                   or Hamiltonians", {\it J. Phys. A: Math. Gen.}, 25, L291 -- L297, 1992. 
\bibitem{38} Z.E. Musielak, N. Davachi and M. Rosario-Franco, "Lagrangians, Gauge Functions, and Lie 
                   Groups for Semigroup of Second-Order Differential Equations", {\it J. Appl. Math.}, 3170130
                   (11 pages), 2020.


\end{thebibliography}
\end{document}